\documentclass[review,5p,times,twocolumn]{elsarticle}

\usepackage{amssymb}

\usepackage{textcomp}

\journal{Nuclear Physics A}

\begin{document}

\begin{frontmatter}


\title{Magnetic Field Devices for Neutron Spin Transport and Manipulation in Precise Neutron Spin Rotation Mesurements}
\author[label1]{M. Maldonado-Vel\'azquez}, \author[label2]{L. Barr\'on-Palos\corref{cor1}}, \author[label3]{C. Crawford}, \author[label4]{W. M. Snow}
 \address[label1]{Posgrado en Ciencias F\'isicas, Universidad Nacional Aut\'onoma de M\'exico, 04510, Mexico}
 \address[label2]{Instituto de F\'isica, Universidad Nacional Aut\'onoma de M\'exico, Apartado Postal 20-364, 01000, Mexico}
 \address[label3]{University of Kentucky, Lexington, Kentucky 40506, USA}
 \address[label4]{Indiana University, Bloomington, Indiana 47405, USA}

 \cortext[cor1]{libertad@fisica.unam.mx}





\begin{abstract}
The neutron spin is a critical degree of freedom for many precision measurements using low-energy neutrons.  Fundamental symmetries and interactions can be studied using polarized neutrons.  Parity-violation (PV) in the hadronic weak interaction and the search for exotic forces that depend on the relative spin and velocity, are two questions of fundamental physics that can be studied via the neutron spin rotations that arise from the interaction of polarized cold neutrons and unpolarized matter.  The Neutron Spin Rotation (NSR) collaboration developed a neutron polarimeter, capable of determining neutron spin rotations of the order of $10^{-7}$ rad per meter of traversed material.  This paper describes two key components of the NSR apparatus, responsible for the transport and manipulation of the spin of the neutrons before and after the target region, which is surrounded by magnetic shielding and where residual magnetic fields need to be below 100 \textmu G.  These magnetic field devices, called input and output coils, provide the magnetic field for adiabatic transport of the neutron spin in the regions outside the magnetic shielding while producing a sharp nonadiabatic transition of the neutron spin when entering/exiting the low-magnetic-field region.  In addition, the coils are self contained, forcing the return magnetic flux into a compact region of space to minimize fringe fields outside.  The design of the input and output coil is based on the magnetic scalar potential method.

\end{abstract}


\end{frontmatter}


\section{Introduction}
The determination of very small neutron spin rotations ($d\phi/dz\approx 1\times 10^{-7}$ rad/m) in the interaction of polarized low-energy neutrons and unpolarized matter is of relevance for different reasons. Corkscrew neutron spin rotations of transversely  polarized neutrons can come from the weak interaction between the neutrons and the atoms of the material they interact with. The nucleon-nucleon (NN) weak interaction remains poorly understood due to the dominance in intensity of the strong interaction and the non-perturbative nature at low energies of the theory that describes it, quantum chromodynamics (QCD). The measurement of the weak coupling constants that appear in the theories that aim to describe the interaction and that cannot yet be calculated accurately \cite{DDH, Zhu, Wasem} is underway by several experimental collaborations. Experiments that involve few-nucleon systems and low-energy polarized neutrons have been developed to measure PV observables that arise from the NN weak interaction \cite{NPDGamma1, NPDGamma2, NSR1}. These observables can be related to the weak coupling constants and a lot of effort has been dedicated to determine enough linearly independent combinations of them, in systems where the uncertainties related to nuclear wavefunctions are smaller than in the case of many-nucleon systems \cite{Ramsey-Musolf-Page, Nico-Snow}.\\ 
A different aspect of neutron spin rotations as a result of the interaction of neutrons with matter is the possibility to search for exotic forces. In particular for the case of forces with relatively long range (\textmu m to cm) that depend on the spin of one (or both) of the interacting particles, experimental constraints are scarce due to the technical challenges that macroscopic amounts of polarized matter impose on precision experiments. However, setting limits on these type of forces is important since a number of theories predict the existence of new interactions mediated by light particles with weak couplings to matter and long ranges, either as a result of broken symmetries in theories beyond the Standard Model, or from theoretical attempts to explain dark matter and dark energy \cite{Jaeckel, Nakamura, Adelberger}.  Polarized low-energy neutrons have emerged as an effective tool to search for these type of forces since intense beams of polarized neutrons are available, their momentum transfer corresponds to the mesoscopic scale and, because of their lack of electric charge, they can pass through macroscopic amounts of matter suffering very small attenuations.\\
The Neutron Spin Rotation (NSR) collaboration has built a slow neutron polarimeter to measure neutron spin rotations that are of the order of tenths of micro radians per meter of  traversed target material. The experimental apparatus, described in detail in \cite{NSR2}, has several purposes: It was first developed to measure the PV neutron spin rotation angle for polarized neutrons passing through unpolarized $^{4}\textnormal{He}$, however it was found that the results of the first stage of this experiment \cite{NSR1} could set the most stringent constraints, in the sub-cm range, for a possible PV exotic force between the neutron and matter, mediated by the exchange of a light spin 1 boson and the first upper bound on parity-odd components of possible in-matter gravitational torsion coupled to neutrons \cite{Yan-Snow}. The upgraded version of the NSR apparatus, to be used to improve the measurement of the PV NSR in n-$^{4}\textnormal{He}$ (weak interaction as well as PV exotic forces) is also being used in an experiment to search for a possible parity-even exotic force between neutrons and matter that depends on the neutron spin and velocity and that is mediated by the exchange of spin 1 light bosons (meV mass range) with axial couplings \cite{NSR3}. This paper describes two key pieces in the NSR apparatus: the input and output coils that transport and manipulate the neutron spin before and after the target low-magnetic-field region. In section \ref{sec:field-requirements} we describe the physics of adiabatic and nonadiabatic neutron spin transport as well as the magnetic field requirements for the NSR apparatus. Section \ref{sec:design} describes the technique, based on the magnetic scalar potential, used in the design of the coils. The characterization of the magnetic fields produced by the coils is described in section \ref{sec:characterization}. Finally, the conclusions are presented in section \ref{sec:conclusion}.

\section{Neutron spin transport and magnetic field requirements for the NSR apparatus}
\label{sec:field-requirements}
\begin{figure}[htb!]
\centering
\includegraphics[width=\columnwidth]{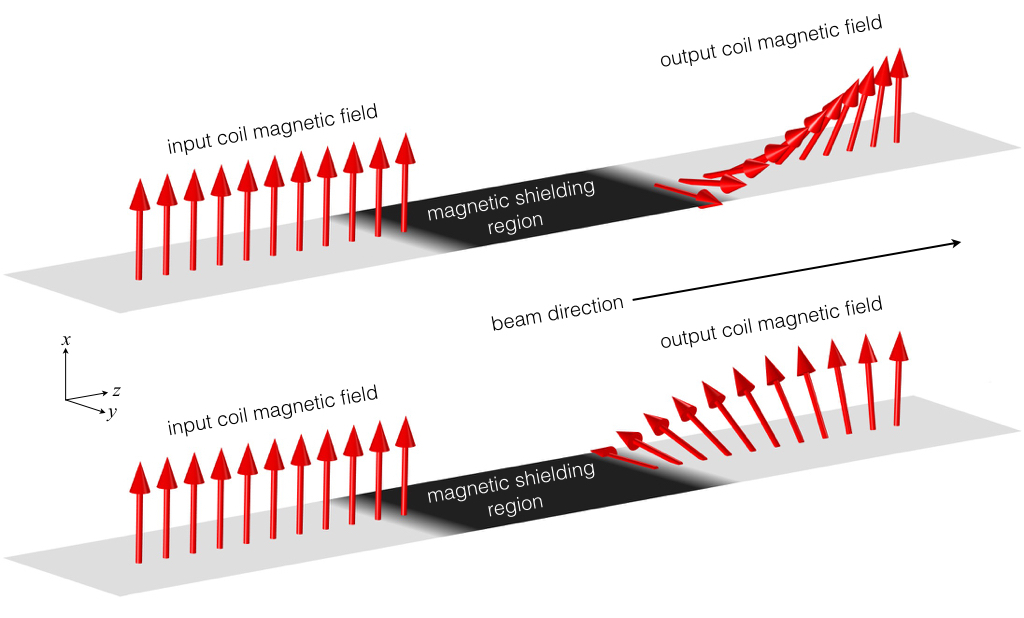}
\caption{Scheme of the magnetic fields for neutron spin transport at the entrance and exit of the magnetic shielding region. The top figure illustrates the magnetic field for counterclockwise rotations of the transverse neutron spin component after the magnetic shielding, while the bottom figure illustrates the field for clockwise rotations.}
\label{fig:IOfields}
\end{figure}

Neutrons are fermions with spin $s=\hbar/2$. They also possess a magnetic moment that is antiparallel to the spin ($\vec{\mu}=-\gamma_{n}\vec{s}$, $\gamma_{n}=1.832 471 72\times 10^{4}$ s$^{-1}$G$^{-1}$). In the presence of an external magnetic field $\vec{B}$, from a semiclassical point of view, a torque is exerted on the neutron spin ($\vec{\tau}=\vec{\mu}\times\vec{B}$), causing it to precess about $\vec{B}$ with a frequency given by $\omega_{L}=\gamma_{n} |\vec{B}|$, known as the Larmor frequency. If the external field is homogeneous, the  projection of the precession axis of the rotating neutron spin will maintain its orientation relative to the external field. In the case of an inhomogeneous external magnetic field, the relative orientation of the neutron spin projection and the field can be maintained if the fractional rate of change of the field as seen in the rest frame of the neutron ($1/\tau=1/|\vec{B}||d\vec{B}/dt|$) is small compared to the precession frequency of the neutron spin ($\omega_{ L}$), or in terms of the adiabaticity parameter $\eta$:

\begin{equation}
\label{eq:adiabaticity}
\eta=\frac{1}{|\vec{B}|}\bigg| \frac{d\vec{B}}{dz}\bigg| \frac{v_{n}}{\omega_L}=\frac{1}{B^{2}}\bigg| \frac{d\vec{B}}{dz}\bigg|\frac{v_{n}}{\gamma_{n}}<<1
\end{equation}

\noindent with $z$ the direction of propagation of the neutron \cite{UCN-book}. On the other hand, if the adiabaticity condition (\ref{eq:adiabaticity}) is not met,  the neutron spin projection will not maintain its relative orientation with the external magnetic field. The ability to establish both adiabatic and non adiabatic conditions is important for neutron spin transport.\\
The NSR apparatus, described in \cite{NSR2}, is a neutron polarimeter that is capable of isolating neutron rotary powers that are in the  $10^{-7}$ rad/m range. The magnetic field of the earth as well as other background magnetic fields present in an experiment can produce much larger neutron spin rotations over distances of order a meter than the neutron-matter interactions of interest. For example, a magnetic field of only 0.5 G (the earth's magnetic field ranges between 0.25 and 0.65 G) can produce a rotation by an angle of 90$^{\circ}$ over a distance of only 5 cm for neutrons with energy of 5 meV (978 m/s of velocity). The determination of very small neutron spin rotations requires, among other things, that the neutron-matter interaction region is surrounded by a magnetic shielding, so that the magnetic fields are below 100 \textmu G. Polarized neutrons, however, need to be transported in and out of this low-magnetic-field region. Entering neutrons require a vertical homogeneous static field to preserve their neutron polarization direction during transport from a super-mirror (SM) neutron polarizer, while exiting neutrons require a static non-homogeneous field to rotate the neutron spin transverse component by 90$^{\circ}$. The transverse component of the neutron spin carries the information of the neutron spin rotation angle and needs to be oriented in the vertical direction to be analyzed by a SM analyzer. The direction of this rotation needs to alternate between counterclockwise and clockwise to reduce systematic effects. The neutron spin transport before and after the target region is adiabatic, however the transitions with the magnetic shielding region has to be nonadiabatic to ensure that neutrons maintain their direction of polarization when no guiding field is present as well as to avoid leakage of magnetic fields into the magnetic shielding region. Two electromagnetic devices, the input and output coils (ic and oc), were developed to produce the magnetic fields for neutron spin transport before and after the magnetic shielding. Figure \ref{fig:IOfields} shows a schematic representation of the magnetic fields.

\noindent The magnetic field produced by the input coil is 
\begin{equation}
\vec{B}_{ic}=B_{o}\hat{x},
\end{equation} 
while the rotating magnetic field of the output coil can be achieved by the combination of a decreasing transverse field and an increasing vertical field,
\begin{equation}
\vec{B}_{oc}= \pm \cos(\pi z/2L)\hat{y}+\sin(\pi z/2L)\hat{x},
\end{equation}
along $L$ the length of the coils. The plus sign corresponds to the counterclockwise rotations of the transverse component of neutron spin while the negative sign corresponds to clockwise rotations. The amplitude of these fields needs to be $\sim$ 4 G in order to comfortably meet the adiabatic condition for cold neutrons. Both coils need to include in their interior the input/output SM neutron guide, radiological shielding and its corresponding holding system, which has a total cross sectional area of 16.5 $\times$ 16.5 cm$^2$. Additionally, the input and output coils need to be self contained, meaning that they have to control the return magnetic flux and keep it in a compact region of space. Finally their cross section needs to be as small as possible to prevent leakage of magnetic fields into the magnetic shielding.

\section{The magnetic scalar potential in the design of coils and the NSR I/O coils}
\label{sec:design}

\begin{figure}[htb!]
\centering
\includegraphics[width=\columnwidth]{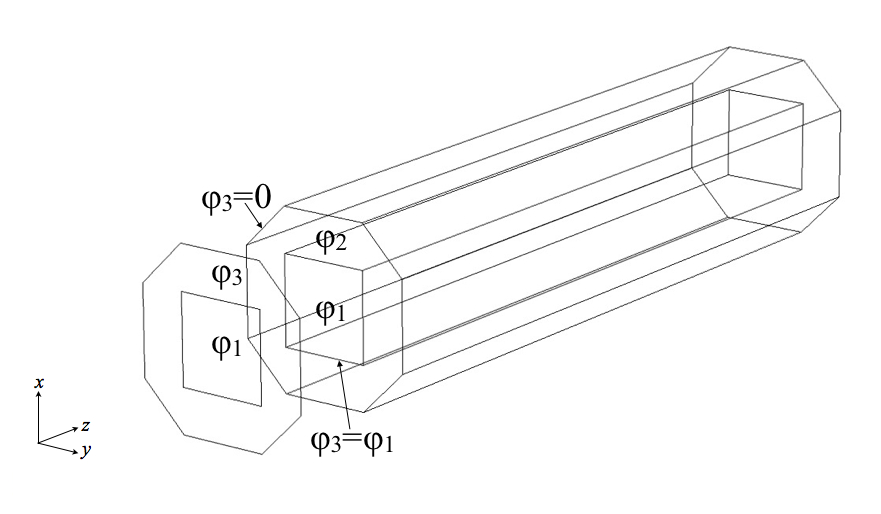}
\caption{Geometry of the input and output coils and their endcap. The different magnetic scalar potential regions are marked. }
\label{fig:geometry-phi}
\end{figure}

The physical significance of the magnetic scalar potential as a `source potential' has practical applications in designing coils to produce precision magnetic fields in situations where the geometry is complicated or the field requirements are difficult to meet using other methods~\cite{Crawford}. Design techniques based on this physical principle have been successfully used in the past few years in the development of various electromagnetic devices for spin transport in experiments in fundamental neutron physics.\\
In a region of space free of currents $\nabla\times \vec{H}=0$, so we can express the magnetic field strength in terms of the magnetic scalar potential $\varphi$ as $\vec{H}=-\nabla \varphi$.  From Gauss's law, $\nabla\cdot\mu\vec{H}=0$, and therefore the magnetic scalar potential must satisfy the Laplace's equation $\nabla^{2}\varphi=0$.  This PDE can be solved as a boundary value problem for the given geometry with appropriate flux boundary conditions on the magnetic field.  The magnetic scalar potential solution forms equipotential surfaces which are perpendicular to the magnetic lines of flux everywhere in space.  The physical interpretation of the magnetic scalar potential is that its isocontours on the boundaries of the given geometry represent the paths of electrical current required to produce the corresponding field in that region.  The current in each path equals the difference in potential between two adjacent isocontours.  For this property to hold, each region of nonzero field must be wound along its scalar potential isocontours.  This geometric interpretation provides a powerful practical method of calculating the windings of a coil based on the desired field and geometry, in contrast with the usual technique of calculating the magnetic field of approximate windings and iterating to a solution.\\        
The input and output coils for the NSR apparatus, as described in the previous section, need to produce static magnetic fields along the path of the neutron beam.  At the interface with the magnetic shielding region, the fields need to abruptly decrease to zero so that a nonadiabatic transition is produced. The coils also need to have an external region to confine the return magnetic flux; this region needs to have approximately the same volume as the internal region (neutron path region) to avoid large fields and achieve good confinement.  The geometry of the input and output coils is shown in figure \ref{fig:geometry-phi}: they are octagonal prisms with a hollow region at the center (neutron path) with 16.5 $\times$ 16.5 cm$^2$ of cross section.  The total width and height of the coils is 37.5 cm and their length along the neutron beam direction is 127 cm.\\
Using the finite element analysis software COMSOL Multiphysics \cite{comsol} to solve the Laplace equation with the magnetic flux boundary conditions described in section \ref{sec:field-requirements}, we calculated the magnetic scalar potential for the internal ($\varphi_1$) and external ($\varphi_2$) regions of the coil, corresponding to a given magnetic flux density $\vec{B}$ and its return flux $-\vec{B}$.  Nominally, these potentials specify two sets of contours, one in each region, which must be wound in order to generate the required fields.  However the inner region shares almost its entire boundary with the outer region, except for the squared surface at each end. Therefore it is more practical to wind both regions as a single coil around the external region, with a small flat endcap coil at one end to correct for current in the missing square (only the end near the magnetic shielding needs a correction endcap since the magnetic field at the other end is continuous with the adjoining guide field from the SM polarizer and analyzer).  In order to shift $\varphi_1$ to the surrounding region $\varphi_2$ and consolidate the two coils, we need to reroute the coil from the inner (beam path) region to the outer region.  We do this by solving an auxiliary Laplace equation for $\varphi_3$, constrained to $\varphi_{3}=\varphi_{1}$ at the interface of the two coils, $\varphi_{3}=0$ on the outer surface, and $\partial\varphi_3/\partial n=0$ on the endcaps (see figure \ref{fig:geometry-phi}).  Now $\varphi_2$ and $\varphi_3$ occupy the same region, and may be added algebraically to produce the isocontours $\varphi_{2}-\varphi_3$ of a single inner/outer winding.\\

\begin{figure}[htb!]
\centering
\includegraphics[width=\columnwidth]{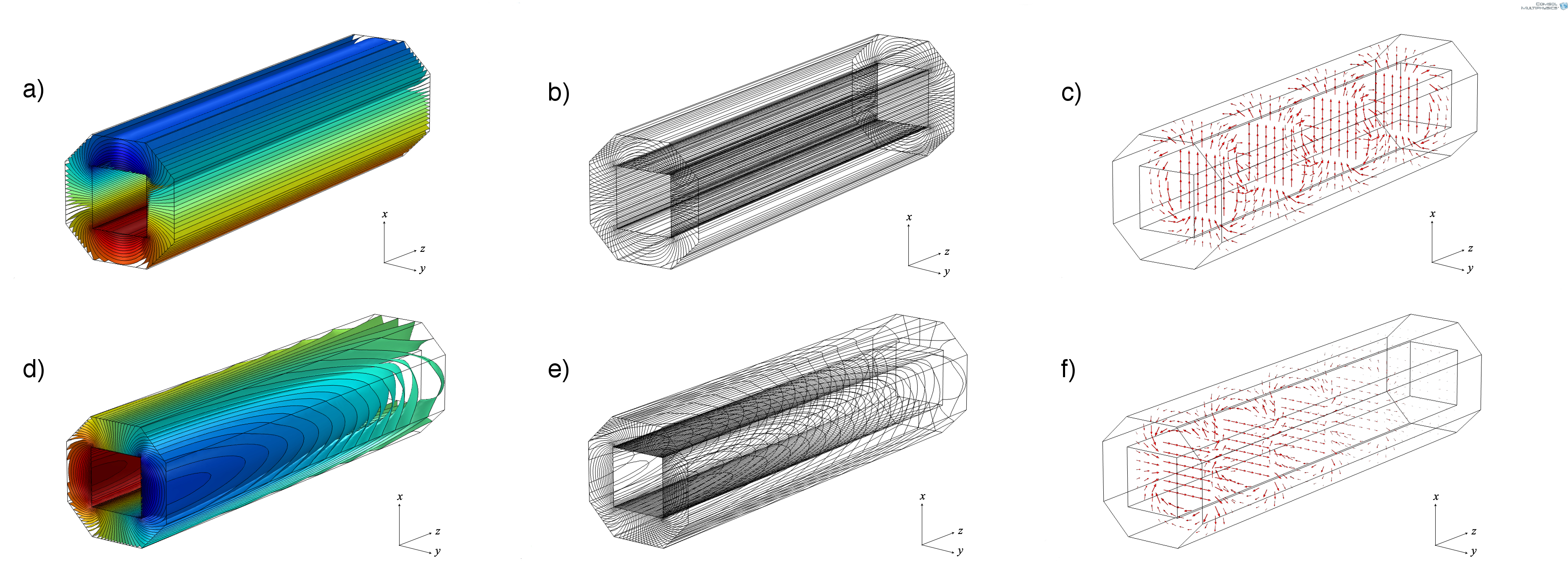}
\caption{a) and d) show the equal magnetic scalar potential for the vertical field of the input coil and the decreasing transverse field of the output coil; b) and e) the corresponding contours of intersection of the isosurfaces and the geometry; c) and f) the magnetic flux.}
\label{fig:phi-Fields}
\end{figure}

\begin{figure}[htb!]
\centering
\includegraphics[width=\columnwidth]{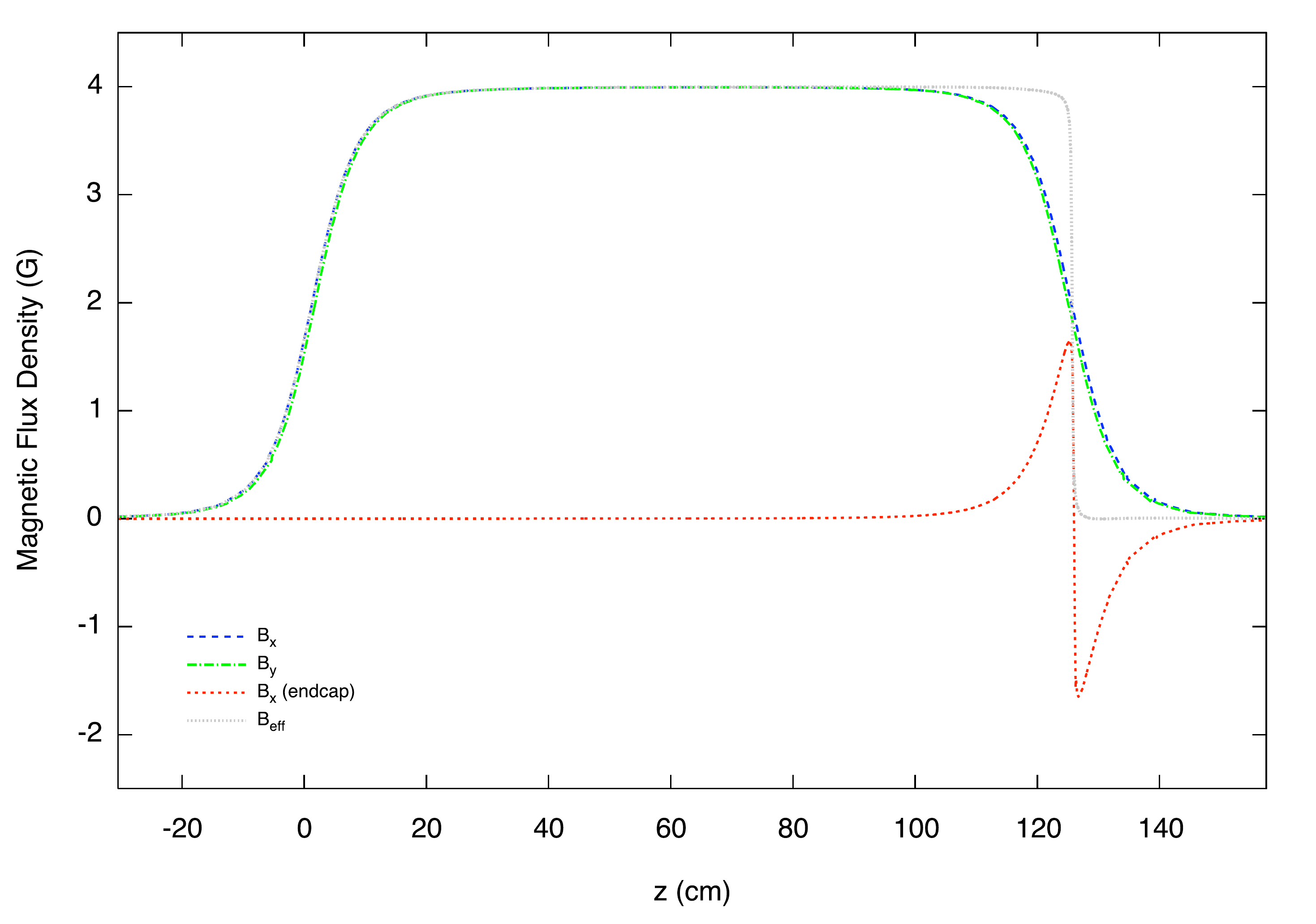}
\caption{Calculated magnetic fields along the axis of the input coil. }
\label{fig:IC-CalculatedFieldsAxis}
\end{figure}

\begin{figure}[htb!]
\centering
\includegraphics[width=\columnwidth]{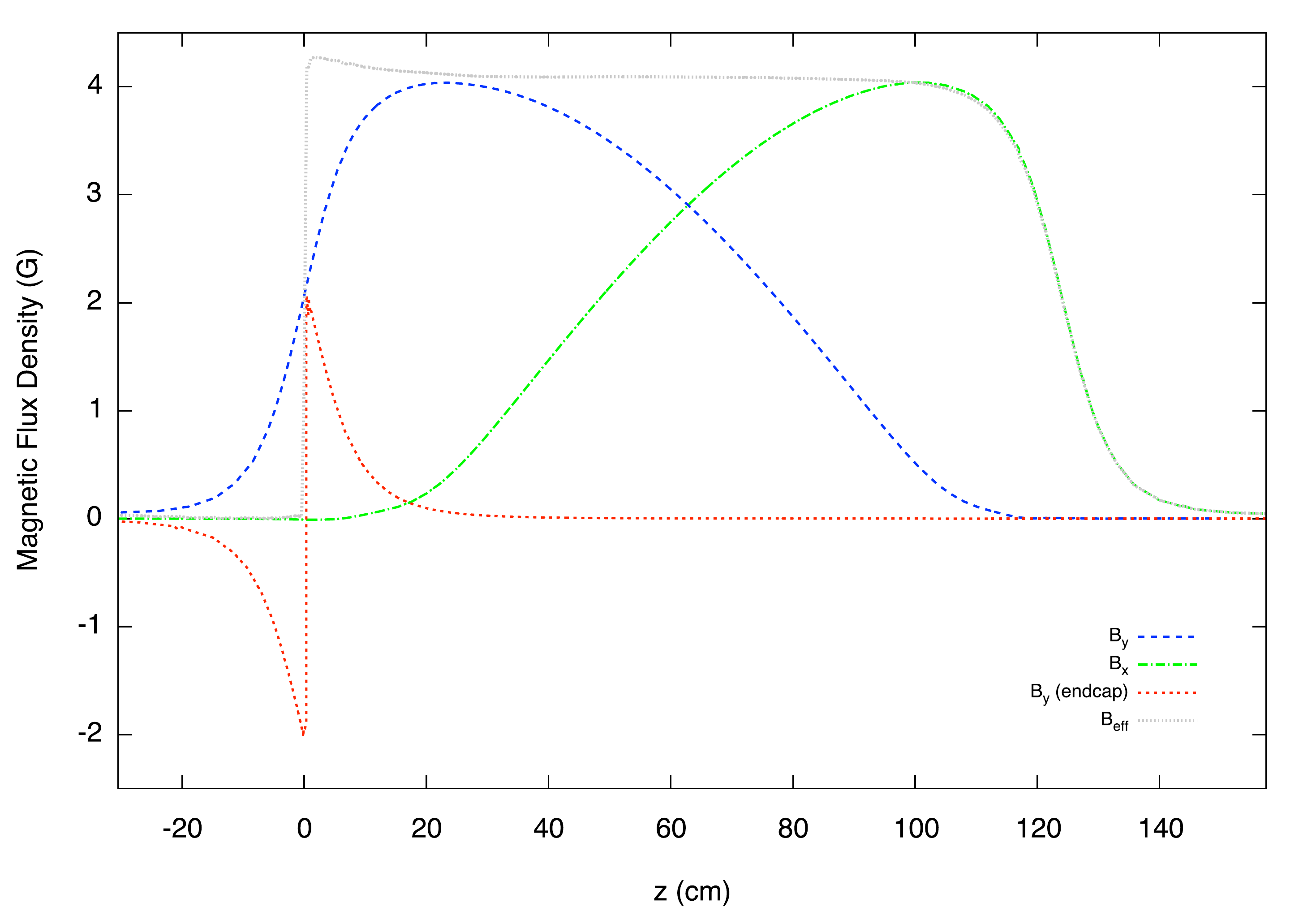}
\caption{Calculated magnetic fields along the axis of the output coil. }
\label{fig:OC-CalculatedFieldsAxis}
\end{figure}

\begin{figure}[h!]
\centering
\includegraphics[width=\columnwidth]{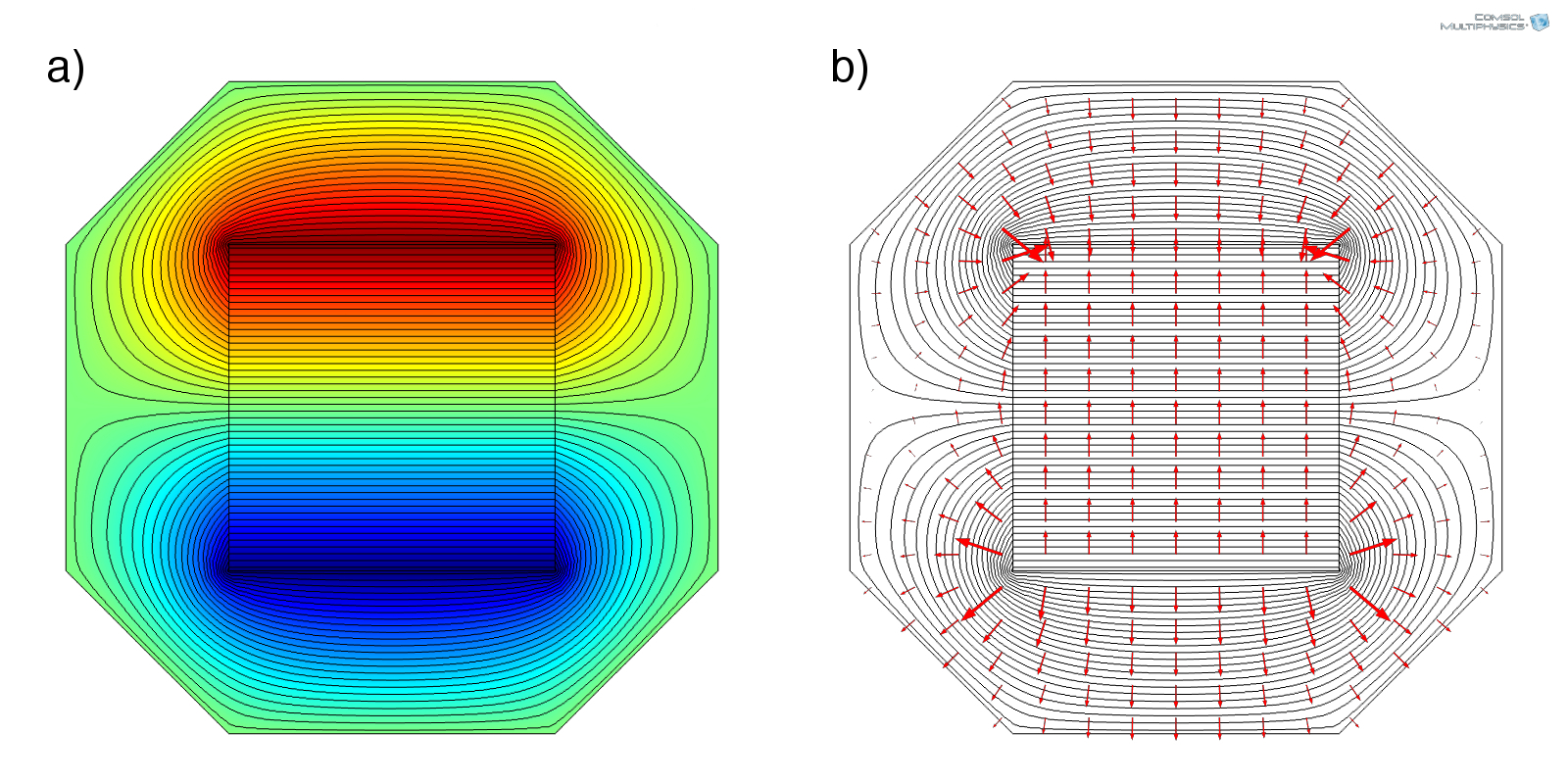}
\caption{Isolevels of scalar magnetic potential on the encap (a) and contours or winding patterns (b).}
\label{fig:endcap}
\end{figure}

\noindent Figure \ref{fig:phi-Fields} shows, for the vertical homogeneous field of the input coil and the decreasing transverse field of the output coil, the equal magnetic scalar potential surfaces ($\varphi_{2}-\varphi_{3}$), the contours at the intersection of the equal potential surfaces and the geometry (winding pattern) and finally the magnetic fields that are generated when electrical current circulates through the contours.  Figures \ref{fig:IC-CalculatedFieldsAxis} and \ref{fig:OC-CalculatedFieldsAxis} show COMSOL calculations of the magnetic fields produced along the axis of the coils.  As can be seen, the generated magnetic field does not have an abrupt step-like profile at the end of the input coil and the beginning of the output coil.  This abrupt transition to zero field is important to ensure nonadiabatic spin transport at the entrance and the exit of the magnetic shielding, and was lost as a result of rerouting currents from the inner $\varphi_1$ region to the outer $\varphi_3$ region. To restore this transition profile of the magnetic fields an endcap for each coil was designed with the contours of $\varphi_1$ in the inside square and $\varphi_3$ in the surrounding octagon.  Figure \ref{fig:geometry-phi} shows the magnetic scalar potentials used for the two regions of the endcap, so that when combined with the main coil we recover the magnetic scalar potentials $\varphi_1$ and $\varphi_2$ in the internal and external regions of the coil at the interface with the target region.  The eqipotential levels, the contours (winding pattern), and the and magnetic fields for the endcaps are shown in figure \ref{fig:endcap}.  The magnetic field produced by the endcaps as well as their combination with the magnetic fields of the coils along the axis are also shown in figures \ref{fig:IC-CalculatedFieldsAxis} and \ref{fig:OC-CalculatedFieldsAxis}. \\

\begin{figure}[h!]
\centering
\includegraphics[width=\columnwidth]{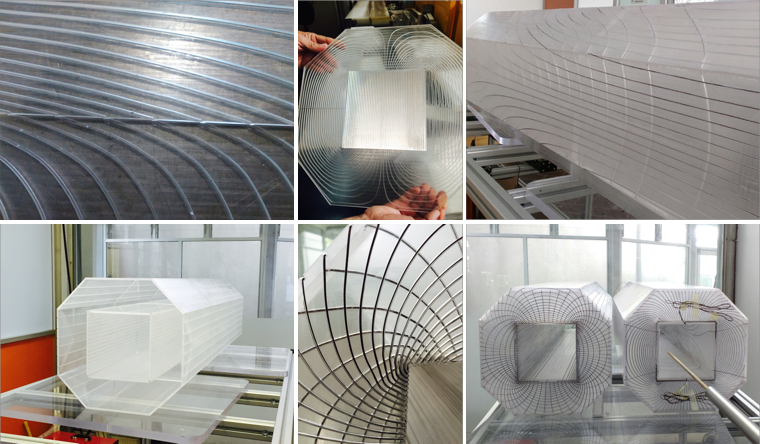}
\caption{Photos of the coils at different stages of construction. Two layers of 5 mm thick acrylic plates were used for the structure of the coils (one layer for each field component); gauge 18 enameled magnet aluminum wire was used to wind the coils; 5 mm thick aluminum plates were used for the structure of the endcaps in the region of the neutron beam path.}
\label{fig:photos}
\end{figure}

\noindent The output coil needs to have two layers of windings: one to produce the transverse decreasing field, which also needs to alternate polarity to produce clockwise or counterclockwise rotations of spin; and the second to produce a vertical increasing field.  The input coil also needs to produce a second magnetic field which is homogeneous and perpendicular to the transport field. This field is used to generate controlled neutron spin rotations that are required to calibrate the apparatus.  For the construction of the coils it is very important that nonmagnetic materials are used.  Because of the low intensity of the magnetic fields ($\sim$ 4 G), the dissipated power for each coil is of the order of 40 W, therefore we were able to use acrylic plates, 5 mm thick, for the structure of the coils.  Channels of approximately 1.3 $\times$ 1.3 mm$^2$ cross section were machined into the surface of the acrylic following the trajectories of the magnetic scalar potential isocontours.  Since the endcaps have current paths crossing the neutron beam trajectory, it was important to use a material that has a low neutron absorption cross section.  Aluminum enameled magnet wire (2 coatings of insulation, modified polyester and polyamideimide), gauge 18, was used to wind the coils placing the wire inside the channels. The required electrical current was $\sim$ 2.5 A, well below the maximum current of 10 A for this type of wire.  For the endcaps, in order to provide a frame for the wires crossing the beam path, 5 mm-thick aluminum plates were used instead of acrylic. Photos of the machined channels on the acrylic, the structure as well as of the wound coils are shown in figure \ref{fig:photos}.  Both coils have two layers of acrylic, one for each field component (vertical and transverse).

\section{Characterization of the coils} 
\label{sec:characterization}

\begin{figure}[h!]
\centering
\includegraphics[width=\columnwidth]{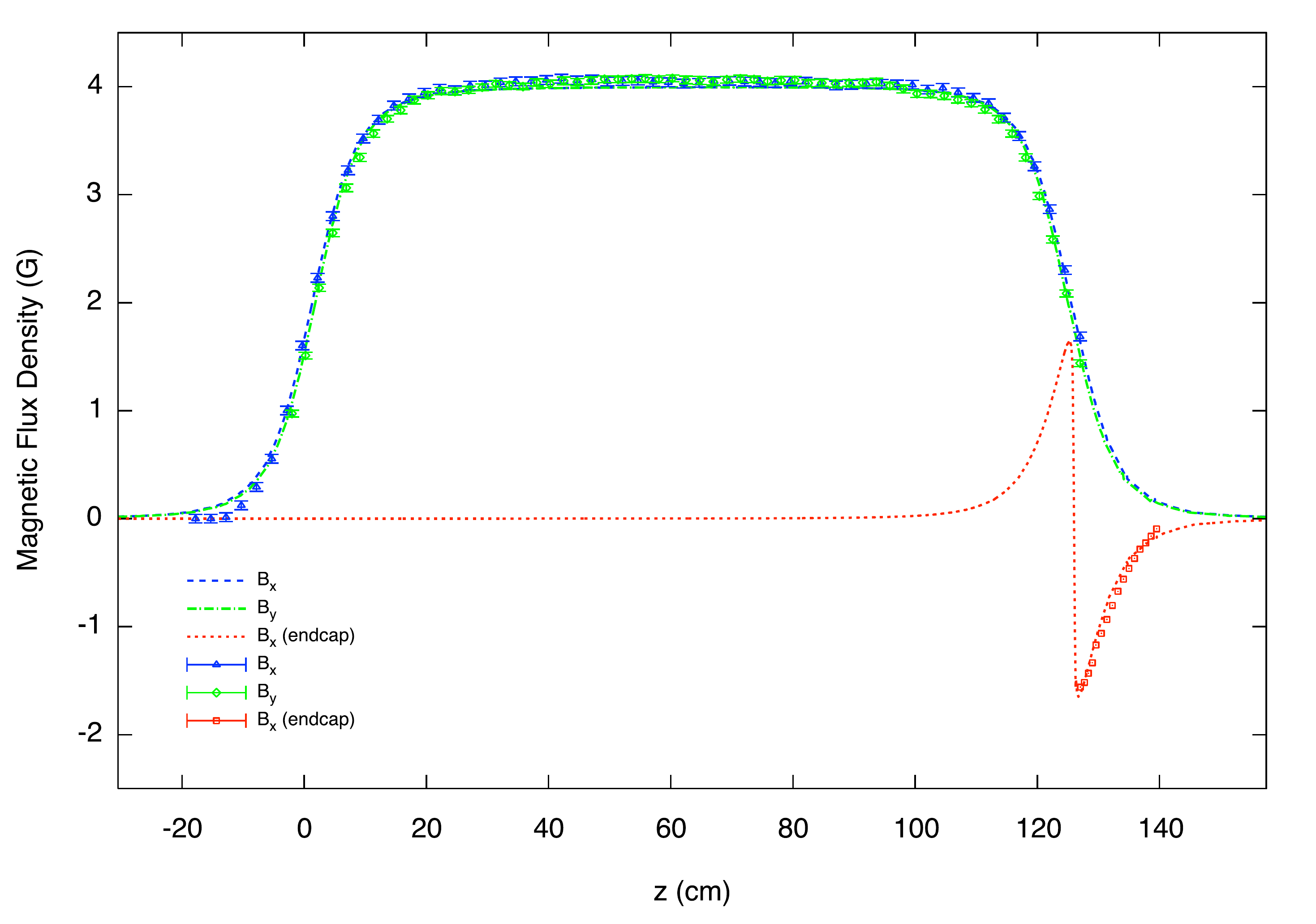}
\caption{Input coil magnetic fields calculated along the axis of the coil. }
\label{fig:ICFieldsAxis}
\end{figure}

\begin{figure}[ht!]
\centering
\includegraphics[width=\columnwidth]{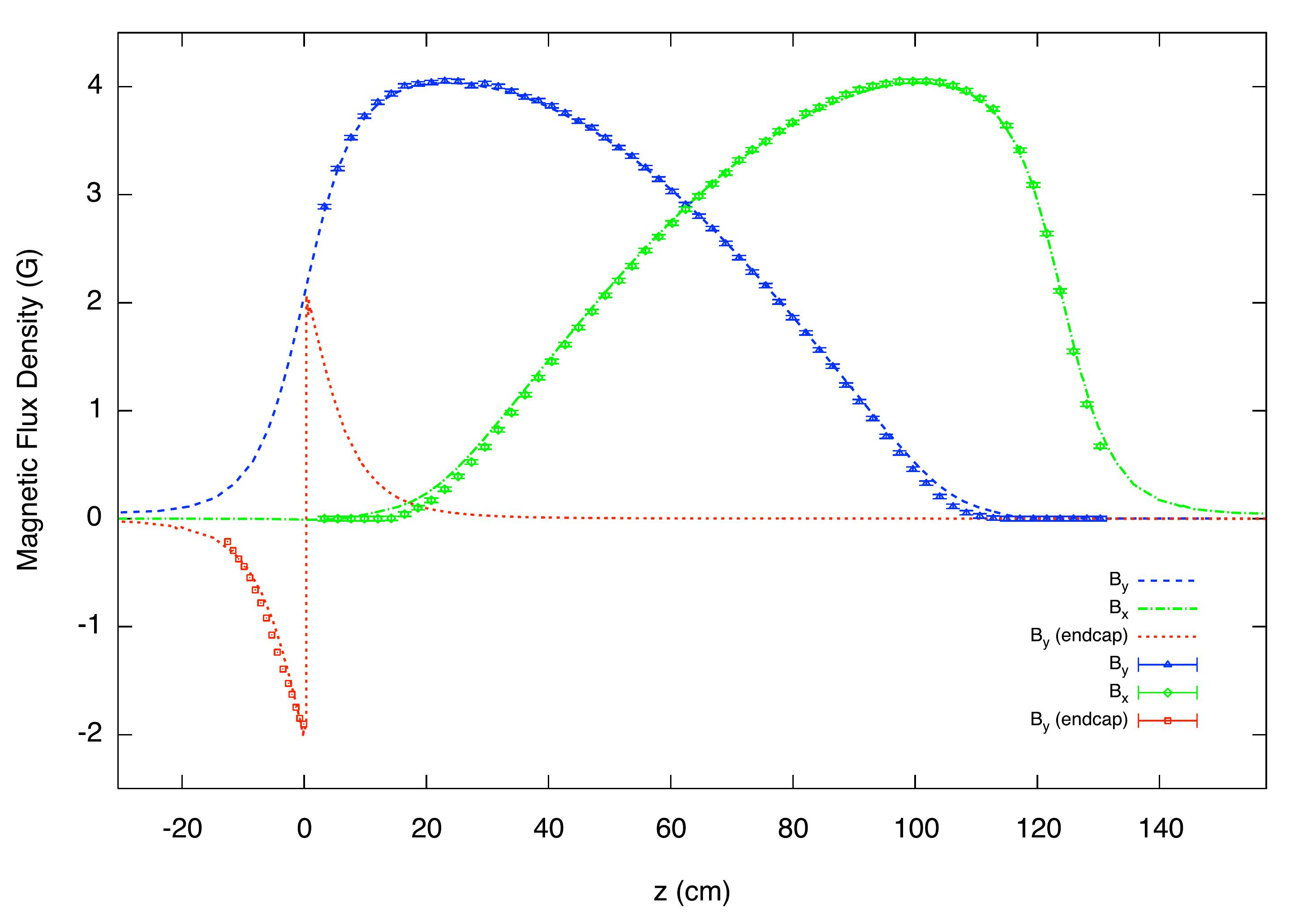}
\caption{Output coil magnetic fields calculated along the axis of the coil. }
\label{fig:OCFieldsAxis}
\end{figure}

Magnetic field maps of the coils were performed using a 3-channel gaussmeter (Lakeshore 460) and a 3-axis probe (Lakeshore MMZ) mounted on a XYZ system consisting of three non-magnetic slides (Velmex MT10 and MN10). The spatial resolution of each slide is 6.35 \textmu m (400 step/rev, equivalent to 0.1 inch advance). To eliminate the background magnetic fields in the field map, a pulse generator was used to energize the coils, using a sinusoidal wave with amplitude of 10 V peak-to-peak and frequency of 200 Hz. Using the RMS setting, the gaussmeter can measure time oscillating magnetic fields with frequencies up to 400 Hz with resolution of 0.001 G. Figure \ref{fig:ICFieldsAxis} and \ref{fig:OCFieldsAxis} show a comparison of the magnetic fields calculated with COMSOL Multiphysics from the design and the measured magnetic fields along the axis of the coil. 

\begin{figure}[h!]
\centering
\includegraphics[width=\columnwidth]{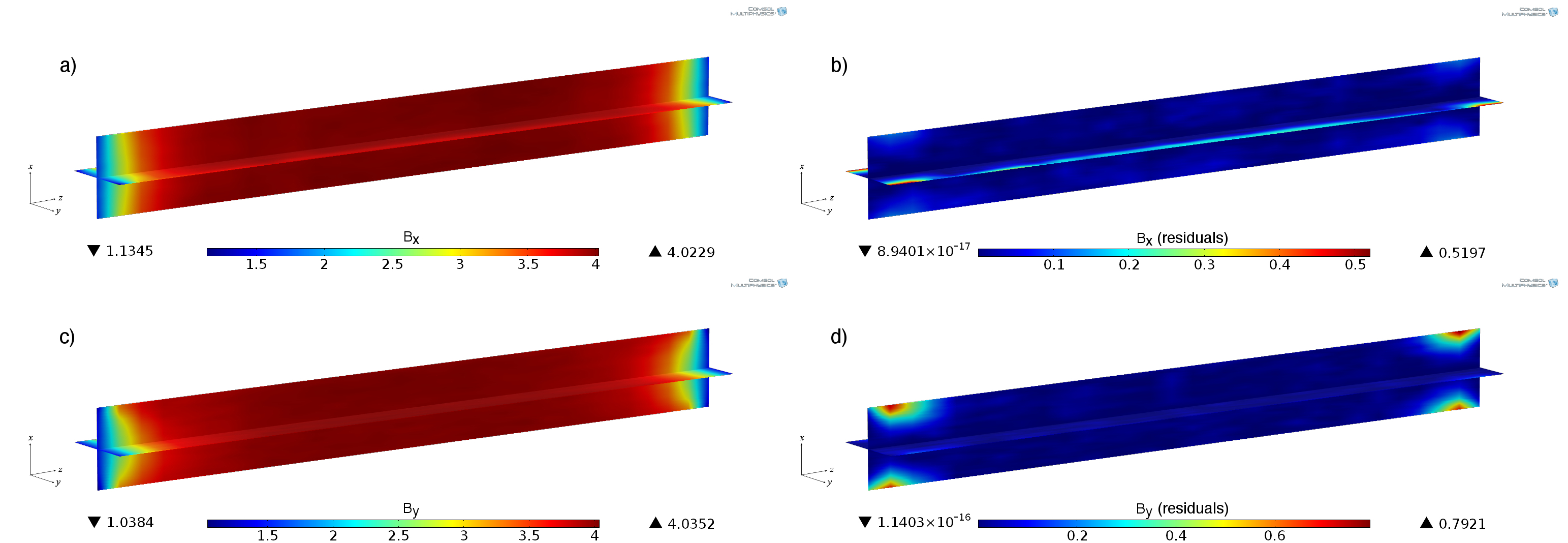}
\caption{Input coil field map and field uniformity. a) and c) show the field map of the vertical field (holding field for polarized neutron beam) and the transverse field (field for calibration of the apparatus) in the volume inside the coil; b) and d) show the residuals of the magnetic field and the magnetic field along the axis of the coil for both components. Neutron beam propagates along the z-axis. }
\label{fig:ICFieldMap}
\end{figure}

\begin{figure}[h!]
\centering
\includegraphics[width=\columnwidth]{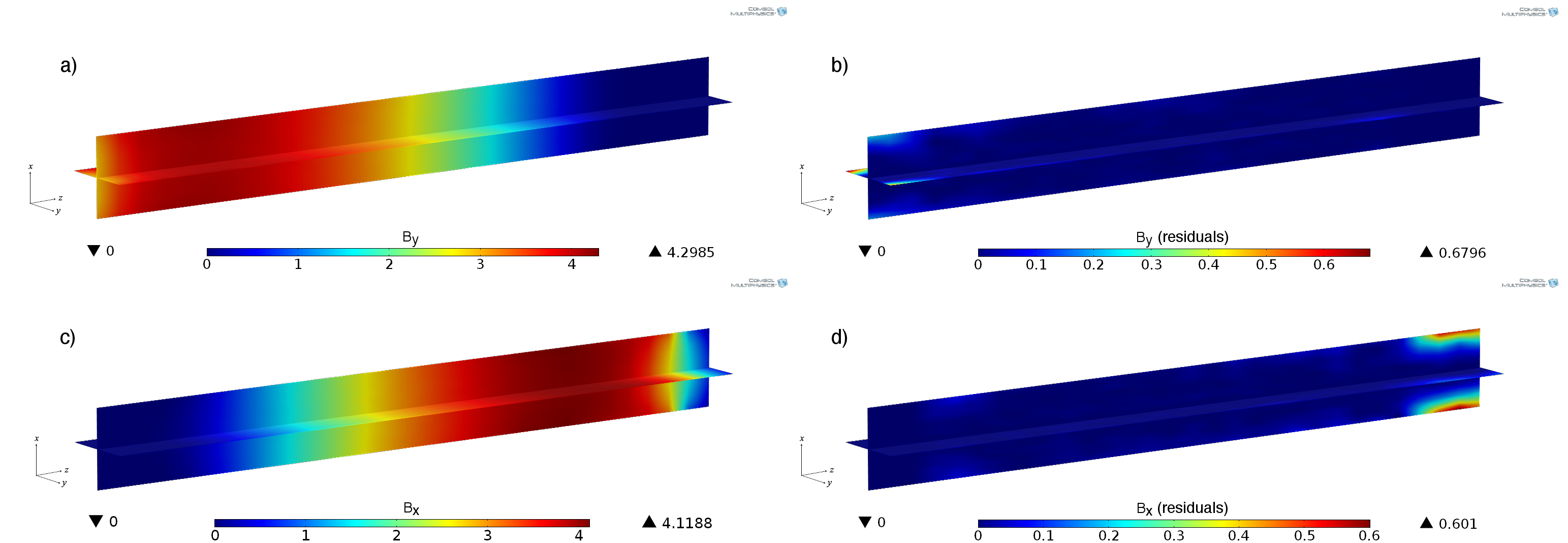}
\caption{Output coil field map and field uniformity. a) and c) show the field map of the two components of the effective magnetic field in the volume inside the coil (transverse and vertical correspondingly); b) and d) show the residuals of the magnetic field and the magnetic field along the axis of the coil for both components. Neutron beam propagates along the z-axis. }
\label{fig:OCFieldMap}
\end{figure}

\noindent The field maps and field uniformity in the internal region of the coils (volume of 14 $\times$ 14 $\times$ 127 cm$^3$) are presented in figure  \ref{fig:ICFieldMap} for the input coil and figure \ref{fig:OCFieldMap} for the output coil; a) and c) show the field map of the vertical and transverse components of the field, while b) and d) show the field uniformity as the residuals of the field and the field in the axis of the coil. The residuals of the magnetic field components have average values over the mapped volume of 0.03 (transverse) and 0.04 (vertical) for the input coil and of 0.02 G for the two components of the output coil. These averages have contributions of larger deviations (up to 0.7 G) from the on axis field that occur near the winding wires, at the edges of the internal region of the coil, however for most of the volume the residuals are below 0.01 G for the two components in both coils. Additionally the inner cross section of the SM input and output guides inside the coils is 10 $\times$ 10 cm$^2$ and therefore the neutron trajectories are away from the edges of the coil. The $z$ component of the fields in all cases is below 0.01 G. 

\begin{figure}[h!]
\centering
\includegraphics[width=\columnwidth]{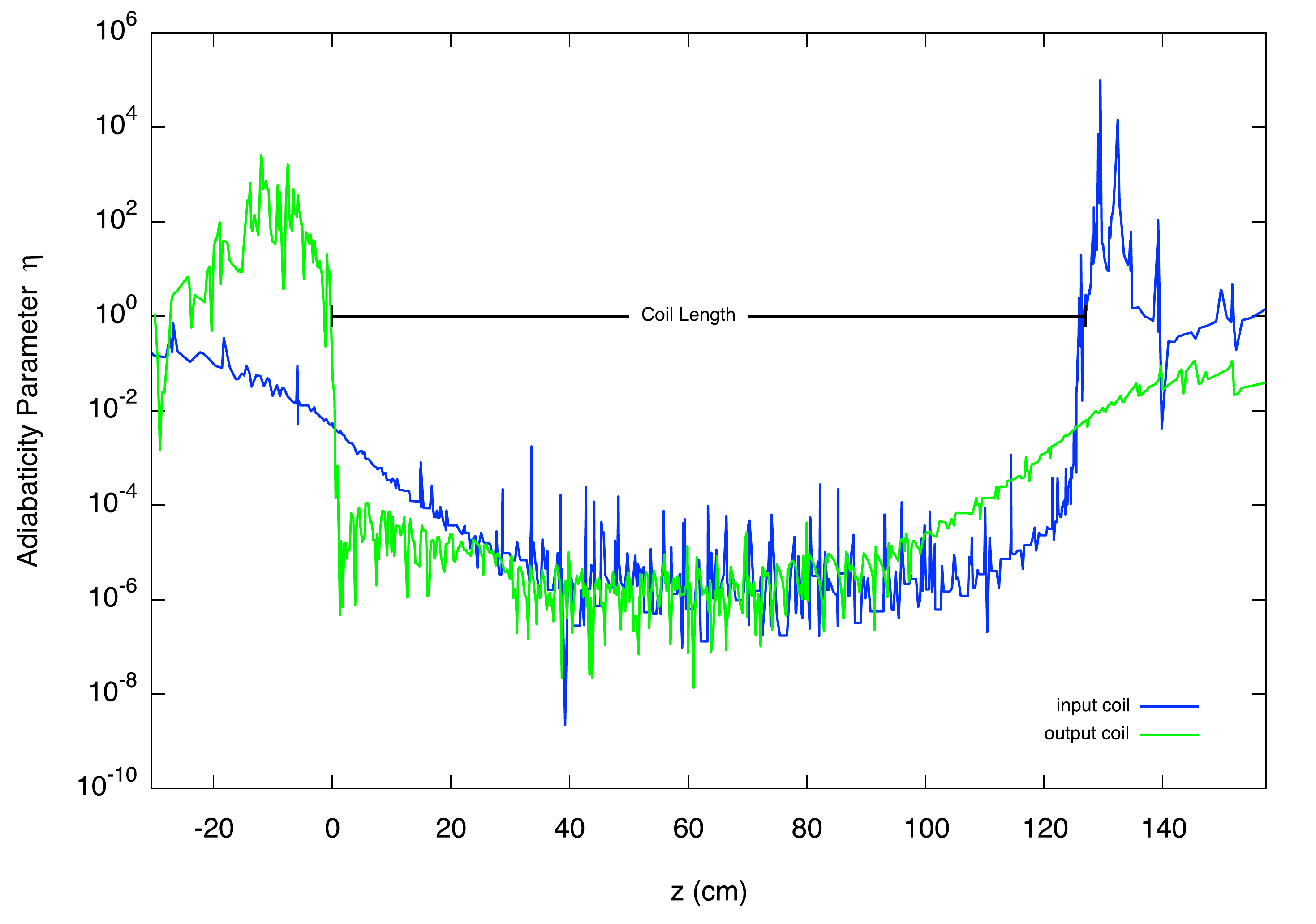}
\caption{Adiabaticity parameter $\eta$ for the input and output coil. Neutron spin transport is adiabatic along the length of the coils ($\eta<<1$), however a non adiabatic transition can be observed at the exit of the input coil and the entrance of the output coil, where the coils interface with the magnetic shielding region.}
\label{fig:Adiabaticity}
\end{figure}

\noindent Finally the adiabaticity parameter $\eta$, described in section \ref{sec:field-requirements}, is calculated for the measured magnetic fields along the axis of the coils; figure \ref{fig:Adiabaticity} shows $\eta$ for 5 meV neutrons for both input and output coils. As required, the neutron spin transport before (input coil) and after (output coil) the magnetic shielding is adiabatic ($\eta<<1$), however the transition with the low-magnetic-field region (exit of the input coil and entrance of the output coil) is non adiabatic.

\section{Conclusions}
\label{sec:conclusion}
The magnetic scalar potential method has been successfully applied in the design of precision coils to produce magnetic fields for slow neutron spin transport in the Neutron Spin Rotation apparatus. The input and output coils, developed with this method, are two electromagnetic devices that provide static and extremely uniform magnetic fields (residuals below 0.01 G) to adiabatically transport the neutron spin between the polarizer and the target, and between the target and the analyzer. The input coil produces a homogeneous magnetic field in the vertical direction (neutron polarization direction) while the output coil produces a field that in the rest frame of the neutrons rotates by 90$^{\circ}$ about the beam direction. Both coils have an abrupt step like transition with the virtually zero magnetic field region where the target is immersed. The coils also confine the return magnetic flux into a small region in space, reducing the leakage of magnetic fields into the target region. These devices were recently used with success in an experiment to search for a possible parity-even exotic force between the neutrons and matter, and will be used in the future in an experiment to measure the PV rotary power of polarized neutrons in $^4$He.

\section*{Acknowledgements}
L. Barr\'on-Palos and M. Maldonado Vel\'azques gratefully acknowledge the support of PAPIIT-UNAM (Grants IN111913 and IG101016) as well as the assistance provided by Roberto Gleason-Villagr\'an, Marco Antonio Veytia-Vida\~na, Hesiquio Vargas-Hern\'andez and Juan Carlos Morales-Rivera in the construction of the coils. The work of W. M. Snow was supported by US National Science Foundation grant PHY-1306942, by the Indiana University Center for Spacetime Symmetries, and DOE contract DE-SC0008107. 





\end{document}